# Large surface conductance and two-dimensional superconductivity in microstructured crystalline topological insulators

Yangmu Li[1,*], Jie Wu[1], Fernando Camino[2], G. D. Gu[1], Ivan Božović[1], and John M. Tranquada[1]

1 Condensed Matter Physics and Materials Science Division, Brookhaven National Laboratory, Upton, New York 11973, USA

2 Center for Functional Nanomaterials, Brookhaven National Laboratory, Upton, New York 11973, USA

Correspondence to: yangmuli@bnl.gov

**Abstract**

Controllable geometric manipulation via micromachining techniques provides a promising tool for enhancing useful topological electrical responses relevant to future applications such as quantum information science[1-4]. Here we present microdevices fabricated with focused ion beam from indium-doped topological insulator $Pb_{1-x}Sn_xTe$. With device thickness on the order of 1 μm and an extremely large bulk resistivity, we achieve an unprecedented enhancement of the surface contribution to about 30% of the total conductance near room temperature. The surface contribution increases as the temperature is reduced, becoming dominant below approximately 180 K, compared to 30 K in mm-thickness crystals. In addition to the enhanced surface contribution to normal-state transport, we observe the emergence of a two-dimensional superconductivity below 6 K. Measurements of magnetoresistivity at high magnetic fields reveal a weak antilocalization behavior in the normal-state magnetoconductance at low temperature and a variation in the power-law dependence of resistivity on temperature with field. These results demonstrate that interesting electrical response relevant to practical applications can be achieved by suitable engineering of single crystals.

Tunable electrical responses and the emergence of superconductivity in topological insulators have garnered broad interest in both academic and industrial communities due to their potential applications[5,6]. Three-dimensional topological crystalline insulators, with an inverted bulk band gap and spin-momentum-locked metallic surface states protected by crystalline symmetry, represent a novel quantum state of matter[6-9]. Theoretically, such topological crystalline insulators have been predicted for the rocksalt crystal structure[7,8,10]. Subsequent photoemission spectroscopy measurements of $Pb_{1-x}Sn_xTe$ observed Dirac states near high-symmetry reciprocal lattice points[11-14]. $Pb_{1-x}Sn_xTe$ features a topological phase transition with doping $x$ and hosts nontrivial surface states for $x$ larger than ~ $0.35^{12,14}$. By compensating for defects with In substitution, the bulk resistivity of $Pb_{1-x}Sn_xTe$ can be unusually large for intermediate $x^{14,15}$. Surprisingly, a slightly higher In concentration can lead to the emergence of superconductivity[16]. As the resistivity of In-doped $Pb_{1-x}Sn_xTe$ is approximately two orders of magnitude greater than that of the most dilute superconductors (e.g., Ca-doped $SrTiO_3$)[15,17], the nature of superconductivity in $Pb_{1-x}Sn_xTe$ is at odds with conventional Bardeen-Cooper-Schrieffer theory[18].

Focused-ion-beam machining, which offers considerable flexibility in precise control of device geometry, has recently been applied to study topological materials, strongly-correlated materials, and unconventional superconductors[1-4]. Geometric control of quantum devices has proved to be an effective method for manipulating surface and bulk responses[3,14]. In order to increase the surface-to-bulk response ratio and thus facilitate studies of surface electrical responses, we have fabricated microscale devices with an FEI Helios Nanolab 600 Focused Ion Beam/Scanning Electron Microscope DualBeam system. In this letter, we present detailed results on two devices, D1 (84.0 x 23.0 x 0.5 µm$^3$) and D2 (75.0 x 30.3 x 1.8 µm$^3$), fabricated from the same In-doped bulk crystal of $Pb_{1-x}Sn_xTe$ [$(Pb_{1-x}Sn_x)_{1-y}In_yTe$ with $x = 0.4$ and $y = 0.08$]. False-color scanning electron microscopy images of D1 and D2 are presented in Fig. 1 (a) and (b), respectively. To minimize the surface damage, we polished the samples with Ga$^+$ ion beam of selective energy and flux density (30 keV / 2.8 nA, 30 keV / 0.92 nA, 30 keV / 0.28 nA, 16 keV / 0.47 nA, 16 keV / 45 pA, 8 keV / 11 pA in consecutive order) before an in-situ lift-out procedure using a micromanipulated probe. The lift-out samples were placed on SiO$_2$/Si (300 nm SiO$_2$ thickness) wafers with gold pads prefabricated using optical lithography and electron-beam physical vapor deposition (Kurt J. Lesker PVD 75 E-beam evaporator). *In-situ* electron-beam-assisted deposition of platinum was used to connect samples with gold pads and the fabricated devices were annealed at 120°C for 20 minutes to ensure ohmic contacts. Both D1 and D2 were made such that their edges are along equivalent crystalline *a* axes.

Zero-field electrical resistivity of D1 and D2 was measured down to 0.3 K using an Oxford $^3$He vacuum insert system and Janis CPX-HF micro-manipulated probe station [Fig. 1 (c) and (d), respectively]. Both devices are very good insulators with resistivity around 8 x 10$^5$ Ω cm at 10 K, indicating negligible doping effect from Ga ions in the bulk. At the same time, the drop in resistivity below 6 K suggests the emergence of superconducting fluctuations, a state that is not present in the parent crystal[14,16]. As shown in Fig. 1 (e) and (f), the resistivity only reaches zero below 0.4 K. The strongly reduced but finite resistivity below 3 K suggests the presence of two-dimensional (2D) superconducting fluctuations above its phase ordering temperature. The lines through the data points in the insets of Fig. 1 (e) and (f) represent fits to the Halperin-Nelson formula for the resistivity of a phase-fluctuating 2D superconductor[19,20]:

$$\rho(T) = \rho e^{-b/\sqrt{T/T_{BKT}-1}} \qquad (1)$$

where $T_{BKT}$ is the Berezinskii-Kosterlitz-Thouless transition temperature[21,22] at which resistivity becomes zero. $\rho$ and $b$ are other fit parameters. The fits yield $T_{BKT} = 0.36$ K, $\rho = 2.98$ x $10^4$ Ω cm, $b = 1.4$ for D1 and $T_{BKT} = 0.38$ K, $\rho = 7.74$ x $10^4$ Ω cm, $b = 1.2$ for D2. The similarities in $T_{BKT}$ and $b$ indicate consistent behavior in both devices. A likely explanation for the 2D superconductivity is that the polishing of the sample surfaces resulted in the insertion of Ga ions (chemically similar to the In dopants) at sufficient density to dope the near-surface region to a density consistent with superconductivity. The expected penetration depth of Ga for the voltages used here is on the order of 10 nm[1], which lead to a surface resistivity of 1192 Ω cm and 860 Ω cm for D1 and D2, respectively, considering two parallel surfaces for each device.

Next we consider the insulating behavior in the normal state. The temperature dependence of resistivity is very useful for distinguishing the surface and bulk contributions. Assuming that the surface and bulk states contribute to conduction channels in parallel, the normalized resistivity ratio, $r(T) \equiv \rho(T)/\rho(T_0)$, can be expressed as[14,23,24]:

$$1/r(T) = 1/r_s + 1/r_b e^{-\Delta/k_B T} \qquad (2)$$

where $T_0 = 300$ K for D1 and 280 K for D2. $1/r_s$ denotes the surface conduction assumed to be temperature independent[14,23,24], and $1/r_b e^{-\Delta/k_B T}$ denotes the thermally activated bulk conductance. The fitted results are shown in Fig 2, with corresponding parameters $r_s = 3.26$, $r_b = 0.39$, and $\Delta = 32.4$ meV for D1 and $r_s = 5.27$, $r_b = 0.24$, and $\Delta = 36.7$ meV for D2. The similarities of $r_b$ and $\Delta$ indicate consistent bulk response in the two devices. The fraction of surface and bulk contributions to total conductance can be evaluated as $r(T)/r_s$ and $r(T)/r_b$, respectively. In Fig. 2(c), we compare the conductance contributions. The temperature at which surface states constitute half of the total conductance ($T_h$) is found to be approximately 30 K for samples with thicknesses between 0.20 and 0.60 mm[14]. The relatively small surface contribution and low $T_h$ limits the sensitivity to the surface state. For D1 and D2, which have a device thickness of 0.5 µm and 1.8 µm respectively, the $T_h$ is enhanced to 182 K and 137 K. The relative surface contributions near room temperature are 30% and 19%, respectively.

The impact of magnetic field on the resistivity of D1 was probed using the 35 T resistive magnet and a ³He cryostat at the DC Field Facility, National High Magnetic Field Laboratory. The measurements were performed with Signal Recovery 7265 lock-in amplifiers and Lake Shore 372 AC resistance bridge systems. The raw magnetoresistivity results are plotted in Fig. 3(a). The normalized magnetoresistivity, $\Delta\rho/\rho_0$, where $\Delta\rho \equiv \rho(\mu_0 H) - \rho_0$, is shown in Fig. 3(b). Above 6 K, electrical resistivity of D1 is finite for the entire field range and we define $\rho_0 \equiv \rho(\mu_0 H = 0\ \text{T})$. Below 6 K, superconductivity exists at low magnetic field and $\rho_0$ is obtained by extrapolating high magnetic field ($\mu_0 H \geq 10$ T) resistivity to zero field, using the form $\rho(\mu_0 H) = \rho_0 + aH^n$. Remarkably, the high-magnetic-field dependence of $\rho(\mu_0 H)$ is described by $n \approx 1$ below 2 K. This field dependence was previously reported in other topological insulator thin films[25,26], and it was associated with the topological surface states. $n$ gradually increases to 1.7 at 20 K, above which it saturates, as shown in Fig. 3(c). Correspondingly, $\Delta\rho(\mu_0 H = 35\ \text{T})/\rho_0$, a measure of the fractional change in magnetoresisity decreases dramatically with temperature.

To further analyze the surface contribution, we convert the magnetoresistivity into magnetoconductance. The magnetoconductance of topological crystalline insulators has been associated with quantum interference between scattering trajectories of surface states [5-9]. With a π Berry phase, destructive quantum interference suppresses back scattering. This destructive quantum interference can be destroyed by applying external magenetic fields, leading to a decrease in the magnetoconductance[14,27]. Magnetoconductance near the superconducting transition temperature is plotted in Fig. 3(d), which can be described by the weak antilocalization behavior using the Hikami-Larkin-Nagaoka equation[27]:

$$\Delta\sigma = \alpha \frac{e^2}{\pi h}[\ln(H_\phi/H) - \psi(H_\phi/H + 1/2)] \qquad (3)$$

where $\psi$ is the digamma function and $H_\phi$ is the phase-coherence characteristic field. For Indium-doped $Pb_{1-x}Sn_xTe$, Fermi surface crosses four Dirac cones and α is predicted to be close to 2. The fits to 4.3 and 10 K data yield $\alpha = 1.23$, $l_\phi = 27.6$ nm and $\alpha = 0.94$, $l_\phi = 15.8$ nm, respectively. $l_\phi = \sqrt{\hbar/4eH_\phi}$ denotes the electron coherence length, which mildly reduces as thermal fluctuations increase, consistent with previously reported results[14].

In summary, we report an unprecedented enhancement of surface conductance for $Pb_{1-x}Sn_xTe$ devices using microstructure techniques. The temperature $T_h$ at which surface states constitute half of the total conductance is increased to about 180 K and a 30% surface contribution is observed near room temperature. Furthermore, we observe an approximately linear magnetoresistivity and the emergence of a two-dimensional superconductivity at low temperatures. These electrical responses, likely related to an enhanced surface conductance, point to a useful engineering of single crystals towards potential applications.

The authors are grateful to E. S. Choi and S. Maier for assistance with the high magnetic field experiment. Work at Brookhaven is supported by the Office of Basic Energy Sciences, Materials Sciences and Engineering Division, U.S. Department of Energy (DOE) under Contract No. DE-SC0012704. This work used resources of the Center for Functional Nanomaterials, which is a U.S. DOE office of Science Facility, at Brookhaven National Laboratory. A portion of this work was performed at the National High Magnetic Field Laboratory, which is supported by National Science Foundation Cooperative Agreement No. DMR-1644779 and the State of Florida.

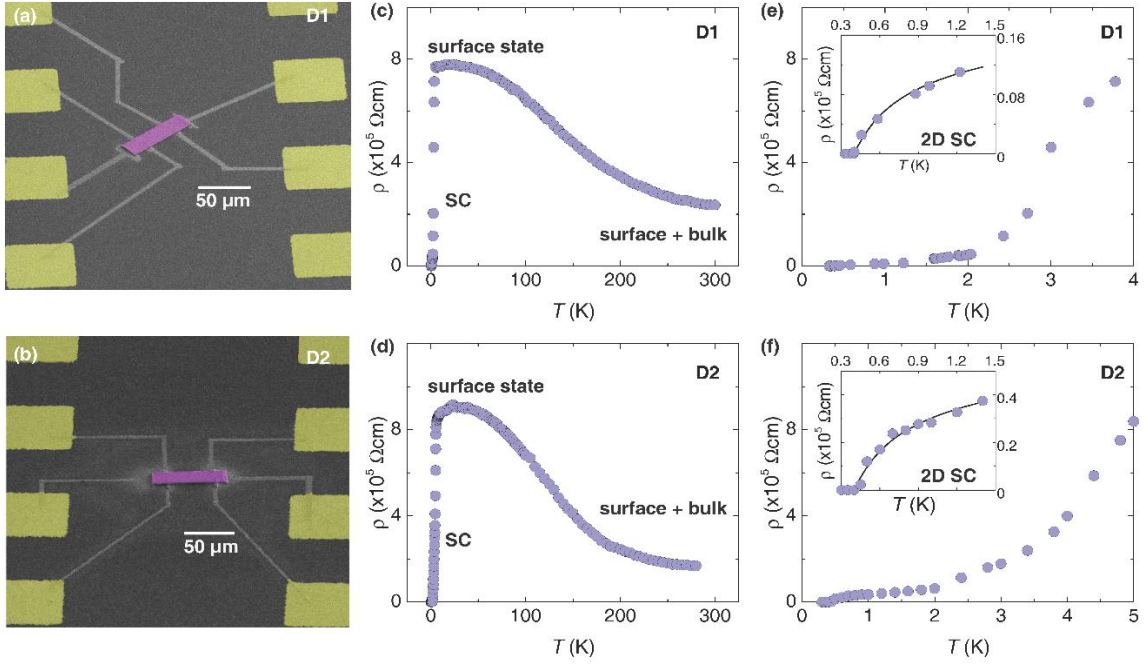

**Figure 1. Pb$_{1-x}$Sn$_x$Te devices and two-dimensional superconductivity. (a, b)** False-color scanning electron microscopy images of Pb$_{1-x}$Sn$_x$Te devices (purple), gold contacts (yellow), and platinum wires (grey) for D1 and D2, respectively. **(c, d)** Corresponding zero-field resistivity. **(e, f)** The emergence of two-dimensional superconductivity. Insets show a fit to Halperin-Nelson 2D SC equation [equation (1)][19,20].

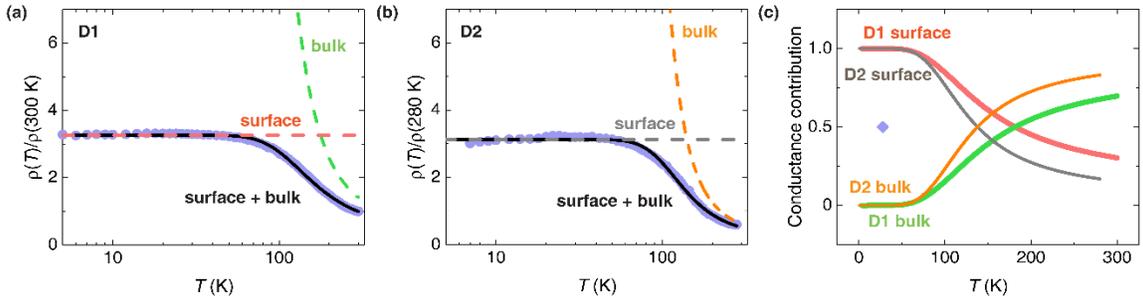

**Figure 2. Surface and bulk contributions. (a, b)** Temperature dependence of normalized resistivity ratio for D1 and D2, respectively. Solid lines denote a fit to parallel conductance equation [equation (2)]. Dashed line present surface and bulk resistivity. **(c)** Temperature dependences of surface and bulk conductance contributions. Blue Dimond indicate $T_h \approx$ 30 K for bulk crystals[14]. $T_h$ = 182 K for D1 and 137 K for D2.

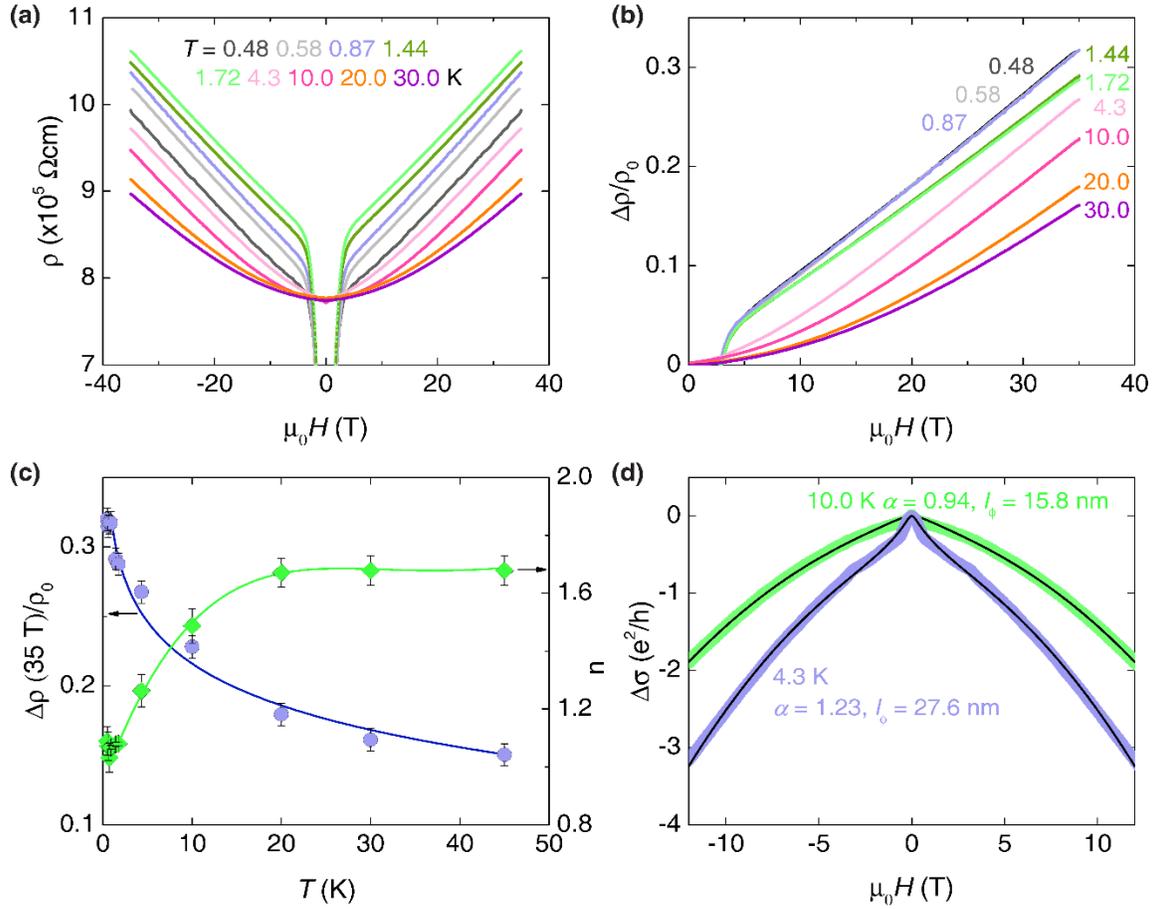

**Figure 3. High-magnetic-field electrical responses for D1.** **(a)** Magnetoresistivity data at various temperatures. **(b)** Normalized magnetoresistivity showing a variation in the power-law dependence of resistivity on temperature with field. **(c)** Temperature dependences of $\rho(35\,\mathrm{T})/\rho_0$ and power index $n$. **(d)** Magnetoconductance and fits to Hikami-Larkin-Nagaoka equation [equation (3)], showing a weak antilocalization behavior[14,27].